\documentclass[aps,pre,twocolumn,superscriptaddress,showpacs]{revtex4}
\usepackage{graphicx}
\begin{document}

\title{Matter wave soliton collisions in the quasi one dimensional
potential}

\author{Nguyen Viet Hung}
\affiliation{Soltan Institute for Nuclear Studies, Ho\.{z}a 69, PL-00-681
Warsaw, Poland}
\author{Micha{\l} Matuszewski}
\affiliation{Nonlinear Physics Center and ARC Center of Excellence for
Quantum Atom Optics, Research School of Physical Sciences and
Engineering, Australian National University, Canberra ACT 0200,
Australia}
\author{Marek Trippenbach}
\affiliation{Soltan Institute for Nuclear Studies, Ho\.{z}a 69, PL-00-681
Warsaw, Poland}
\affiliation{Institute for Theoretical Physics, Warsaw
University, Ho\.{z}a 69, PL-00-681 Warsaw}

\begin{abstract}
We consider soliton solutions of a two-dimensional nonlinear system with the self-focusing nonlinearity and a quasi-1D confining potential, taking harmonic potential as an example. We investigate a single soliton in detail and find criterion for possible collapse. This information is then used to investigate the dynamics of the two soliton collision. In this dynamics we identify three regimes according to the relation between nonlinear interaction and the excitation energy: elastic collision, excitation and collapse regime.  We show that surprisingly accurate predictions can be obtained from variational analysis.
\end{abstract}

\maketitle

\section{Introduction}

Solitons are universal entities in the nonlinear science and interactions between them are perhaps the most fascinating features. Understanding soliton collisions is both of fundamental interest and of importance to its wealth and proposed applications. If the system is integrable, the collisions are elastic \cite{Zahar}. An example of such a system is 1D soliton in attractive nonlinear Kerr medium. However, even if we weakly perturb nonlinear Schr\"odinger equation, we can observe a chaotic character of two-soliton collisions \cite{Dimitriev,Gardiner}. In the general case, it is necessary to resort to detailed numerical calculations for predictions. Stegeman and Segev \cite{SSSience} have introduced a general classification of collisions into two categories: coherent and incoherent. Coherent interactions occur when the nonlinear
medium can respond to interference effects that take place when the beams overlap. Incoherent interactions, on the other hand, occur when the relative phase between the soliton varies much faster than the response time of the medium \cite{ALPapier}. In Ref.~\cite{SSSience} and for example in Ref.~\cite{Ciamaro} one can find description of spatial soliton collisions in the Kerr medium. Collisions between spatiotemporal solitons of different dimensionality in a planar waveguide were considered in \cite{Hector}. Here we also study collisions in the Kerr medium, which has infinitely short response time, but we are dealing with a pair of two dimensional solitons, both of the same dimensionality. We investigate head on collision of solitons moving in quasi 1D confining potential. Our numerical results were obtained for the harmonic potential, but they generally apply to all quasi 1D confining potentials that posses at least one bound state. We would like to point out that multidimensional solitons were extensively studied in the case of Bose Einstein condensates, including beautiful experiments \cite{BECexp}. Some of the applications of solitons in the condensates was discussed in Ref. \cite{Solzat}.

Within the framework of current publication the statics and the dynamics of solitons in quasi 1D potentials are described within the variational approximation and
compared with full numerical simulations. We find formulas for the widths and chemical potential and discuss quasi 1D limit and the dynamics of the collapse. Similar considerations were presented by Salasnich {\it et al} in reference \cite{Salasnich} in the case of Bose-Einstein condensate \cite{collapse_exp,collapse_exp2} and earlier for the optical beam propagation by Li {\it et al} \cite{Li}. Analogous study of the stability of gap solitons were presented in \cite{Obert}. Finally we address the problem of collisions, in various regimes, from elastic collision regime, through the domain where the transverse excitations occur, up to the collapse during the collision. The main result of our study is the analysis of the collapse during the collision. We find it somehow
surprising how good estimate can obtained from a simple variational model. We would like to point out that the deviation from one dimensionality in stationary properties and collisional dynamics of matter-wave solitons was recently investigated in \cite{Kajak} using an effective one-dimensional Gross-Pitaevskii equation that includes an additional quintic self-focusing term. 2D soliton collisions were also studied experimentally in the case of spin wave envelope solitons in Ref.~\cite{Buttner}.

\section{Quasi-1D approximation}

\begin{figure}[tbp]
\includegraphics[width=6cm]{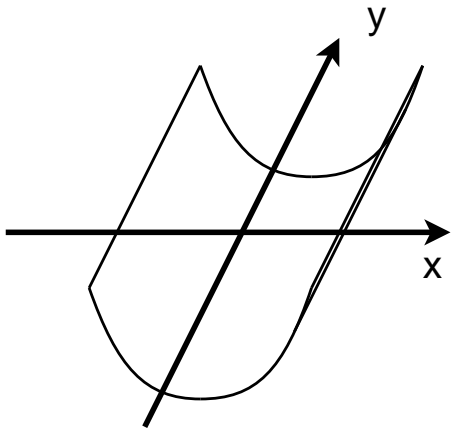}
\caption{Quasi 1D potential with harmonic confinement.} \label{potenfig}
\end{figure}

In this paper we consider two dimensional system with attractive nonlinearity, with an additional external quasi-1D potential. This system is described by 2D nonlinear Schr\"odinger equation (NLSE)
\begin{equation}\label{2DGP}
i\Psi_{t}=-\frac{1}{2}(\Psi_{xx}+\Psi_{yy})+V\Psi-\lambda_{2D}|\Psi|^2\Psi,
\end{equation}
where $\lambda_{2D}>0$. We assume that the wavefunction is normalized to  $N$. All the calculations were performed for the specific case quasi 1D harmonic confining potential $V(x,y)=\omega x^2/2$, see Fig.~(\ref{potenfig}). Our predictions and physical properties derived here apply all the systems, two- and three dimensional, which have an external potential,
confining in all but one dimension (in literature they are sometimes called "potentials with transverse confinement") that support at least one bound state in the transverse direction. From the existing literature on that subject we would like to acknowledge three examples that are closely related to our study: 1)Li {\it et al} \cite{Li} considered the
simple variational model of spatial solitons in planar waveguides, 2) The existence of 3D solitons in the inhomogeneous medium with harmonic potential was investigated by Raghavan \cite{Raghavan}, 3) The dynamics of systems under transverse confinement were studied in BEC by Salasnich {\it et al} \cite{Salasnich}.

In the case of BEC, the equation (\ref{2DGP}) can be derived from the full three-dimensional Gross-Pitaevskii equation for the wavefunction in
physical coordinates $\tilde{\Psi}(\tilde{\bf r}, \tilde{\bf t})$
\begin{equation}\label{3DGP}
i\hbar\tilde{\Psi}_{\tilde{t}}=-\frac{\hbar^2}{2m}\tilde{\nabla}^2 \tilde{\Psi}+U({\bf \tilde{r}})\tilde{\Psi}-\lambda_{3D}|\tilde{\Psi}|^2\tilde{\Psi},
\end{equation}
where $\lambda_{3D} = -4 \pi a_s \hbar^2 / m$, $m$ is the atomic mass, wavefunction norm $N$ is the number of atoms, and $a_s$ is the scattering length.
By assuming strong harmonic confinement $\omega_z$ in the $z$ direction, $U({\bf \tilde{r}}) = m\omega_z^2 \tilde{z}^2/2 + U_{\perp}(\tilde{x},\tilde{y})$, that prevents excitation of higher modes of the trap in this direction, we arrive in dimensionless Eq.~(\ref{2DGP}) after performing rescaling according to $(x,y) = (\tilde{x}, \tilde{y}) / x_0$, $t=  (\hbar / m x_0^2) \tilde{t}$, $V = (m x_0^2/\hbar^2) U$, $\lambda_{2D} = (m /\hbar^2) \sqrt{m \omega_z / 2 \pi \hbar} \, \lambda_{3D}$ and $\Psi =  x_0 \tilde{\Psi}$. Here $x_0$ is an arbitrary scaling parameter.

In all the nonlinear systems with quasi 1D potentials, as long as the transverse part of the potential supports at least one bound state, we expect, in certain range of the strength of the nonlinearity, to find 2D soliton solutions. More precisely, in such systems for small values of nonlinearity we find stable solitons, but there is a threshold, a critical value of nonlinearity at which catastrophic self-focusing occurs. When we approach this critical value our soliton turns into the
Townes soliton \cite{Chiao}, and above this value no stable solitons are
available.

As we mentioned above, to focus our attention in what follows we concentrate on the model with harmonic confinement. In the limit when the trapping frequency becomes large we expect that the energy associated with the nonlinear interaction becomes negligible in comparison with $\hbar \omega$.  In this case our 2D soliton will become a product of a ground state in the direction of the confining potential and 1D soliton in an unbound direction. The following calculation, based on the separation of variables, confirms this statement.

If we assume that our wavefunction $\Psi(x,y,t)$ can be presented as a product of
\begin{equation}
\Psi(x,y,t)=\phi(x)\tilde{\psi}(y,t),
\end{equation}
where $\phi(x)$ is normalized to unity, we can rewrite equation (\ref{2DGP})
\begin{eqnarray}\label{2DGP_separ}
[\phi(x)\tilde{\psi}(y,t)]_{t}=\left[-\frac{1}{2}\phi_{xx}(x)+
\frac{1}{2}\omega^2x^2\phi(x)\right] \tilde{\psi}(y,t) \nonumber
\\ -\left[\frac{1}{2}\tilde{\psi}_{yy}(y,t)
+\lambda_{2D}|\tilde{\psi}(y,t)|^2\tilde{\psi}(y,t)|\phi(x)|^2\right]\phi(x),
\end{eqnarray}
and if we assume that $\phi(x)$ is a ground state of the trapping potential we obtain
\begin{equation}
\left[\tilde{\psi}_{t}-\frac{1}{2}\omega^2\tilde{\psi}\right]\phi=-
\frac{1}{2}\tilde{\psi}_{yy}\phi + \lambda_{2D}|\tilde{\psi}|^2\tilde{\psi}|\phi|^2\phi.
\end{equation}
Upon multiplying both sides of the equation by $\phi^*(x)$, integrating over $x$ and neglecting constant term we obtain one dimensional (NLSE) for $\psi=\tilde{\psi}\exp(-\omega^2t/2)$ (see Eq.~(\ref{1DNLS}) below),
with effective nonlinearity equal to
\begin{equation}\label{NL2Dto1D}
\lambda_{1D}=\lambda_{2D}\int dx |\phi(x)|^4.
\end{equation}
We will call this regime a quasi 1D limit. Notice, that in the case of harmonic potential $\int dx |\phi(x)|^4 = \sqrt{\omega/(2\pi)}$.

One dimensional NLSE is fully integrable. Consequently, we expect that two solitons, that are coming into collision will asymptotically reassume their original shape after collision. If the nonlinear energy becomes
comparable with the excitation energy of the confining potential, we expect to see growing deviation from 1D dynamics. It is the main purpose of this presentation to study this deviation.

\section{Variational Approximation}

\subsection{1D soliton}

The nonlinear Schr\"odinger equation in 1D case reads
\begin{equation}\label{1DNLS}
i\psi_{t}=-\frac{1}{2}\psi_{yy}-\lambda_{1D}|\psi|^2\psi,
\end{equation}
where we assume that norm of the wavefunction is equal to $\int_{-\infty}^{\infty}|\psi|^2dy=N$.
Within the variational approximation \cite{Progres,VABEC}, instead of solving Eq.~(\ref{1DNLS}) we introduce Lagrange density
\begin{eqnarray}
{\cal{L}}=\frac{1}{2}
\left[i(\psi_{t}\psi^{*}-\psi_{t}^{*}\psi)-|\psi_{y}|^2+\lambda_{1D}|\psi|^4
\right].
\end{eqnarray}
To mimic the dynamics we will introduce a trial function of the form (variational Ansatz)
\begin{equation}\label{Ansatz}
\psi(y,t)=A(t)\exp\left(-\frac{y^2}{2}
\left[\frac{1}{V(t)^2}-ib(t)\right]+i\phi(t)\right),
\end{equation}
with variational parameters: amplitude $A(t)$, chirp $b(t)$, overall
phase $\phi(t)$ and width $V(t)$. By substituting Ansatz (\ref{Ansatz})
into our Lagrange density and integrating over $y$ we obtain the Lagrange
function $L=\sqrt{\pi}A^2\left[-\dot{\phi} V-\frac{\dot{b} V^3}{4}-\frac{1}{4V}-
\frac{b^2V^3}{4}+\frac{\lambda_{1D}A^2V}{2\sqrt{2}}\right]$.
Our variational approximation restricts the set of available solutions to the certain class of functions. It will lead to the Euler - Lagrange equations of the reduced Lagrangian shown above
\begin{eqnarray}\label{1DOL}
\dot{\phi} &=& \frac{\lambda_{1D}A^2}{\sqrt{2}}-\frac{1}{4 V^2}-\frac{V \ddot{V}}{4}, \nonumber \\
\dot{V} &=& \frac{1}{V^3}-\frac{\lambda_{1D}}{\sqrt{2\pi} V^2} \,\, \mbox{and} \,\,\,\, b = \frac{\dot{V}}{V}
\end{eqnarray}
with additional one, which we can interpret as a conservation law (first integral) $A^2V = const$,
and which is related to the norm of the trial function
\begin{equation}
\int_{-\infty}^{\infty}|\psi|^2dy=A^2 V\sqrt{\pi} = N \,\,\,\Rightarrow\,\,\,\,
A=\sqrt{\frac{N}{V\sqrt{\pi}}}.
\end{equation}
In this formalism solitons correspond to stationary solutions of Eq.~(\ref{1DOL}) i.~e. we assume $\dot{V}=\ddot{V}=0$.  These conditions can be satisfied when $V=\frac{\sqrt{2\pi}}{N \lambda_{1D}}$ and $b=0$. Notice that we can evaluate soliton eigenvalue. In nonlinear optics this eigenvalue corresponds to the soliton wavevector and in the theory of BEC it becomes chemical potential. We write the
solution in the form
\begin{equation}
\psi(y,t)=\Phi(y)\exp\left( -i\mu_{1D}t \right).
\end{equation}
To evaluate $\mu_{1D}$ we substitute $V$ into the equation for phase, and obtain
\begin{equation}
\dot{\phi}= - \mu_{1D} = -\frac{3\lambda_{1D}^2N^2}{8\pi}.
\end{equation}

\subsection{2D soliton}

In this case we start with 2D nonlinear Schr\"odinger equation with harmonic potential
\begin{equation}
i\Psi_{t}=-\frac{1}{2}(\Psi_{xx}+\Psi_{yy})+\frac{1}{2}\omega^2x^2\Psi-\lambda_{2D}|\Psi|^2\Psi,
\end{equation}
with normalization$\int\int_{-\infty}^{\infty}|\Psi|^2dxdy=N$.
The 2D Lagrange density is
\begin{eqnarray}
{\cal{L}}&=&\frac{1}{2}
\left[i(\Psi_{t}\Psi^{*}-\Psi_{t}^{*}\Psi)-|\Psi_{x}|^2-\right.  \nonumber \\
&&\left.-|\Psi_{y}|^2-\omega^2x^2|\Psi|^2+\lambda_{2D}|\Psi|^4 \right].
\end{eqnarray}
We use 2D Gaussian Ansatz
\begin{eqnarray}
\Psi(x,y,t)  &=&
A(t)\exp\left(-\frac{x^2}{2W(t)^2}-\frac{y^2}{2V(t)^2}\right) \times
\nonumber \\
&& \times
\exp\left(i\left[\phi(t)+\frac{1}{2}(b(t)x^2+c(t)y^2)\right]\right),
\nonumber
\end{eqnarray}
with variational parameters $A(t)$, $\phi(t)$, $b(t)$, $c(t)$, $W(t)$ and $V(t)$. In analogy with what we presented above we obtain (after integrating over $x$ and $y$) the Lagrangian
and look for the stationary solutions ($\dot{W}=\ddot{W}=0$ and $\dot{V}=\ddot{V}=0$), which occur when the following conditions are satisfied: $c = 0$, $b=0$ and
\begin{equation}\label{2DSS}
V =  \frac{2\pi}{\lambda_{2D}N}W,\,\,\, \mbox{and}\,\,\,
W = \sqrt[4]{\frac{4\pi^2-\lambda_{2D}^{2}N^2}{4\pi^2\omega^2}}
\end{equation}
First significant observation is the clear evidence of collapse in our model. Note that equation for $W$ can not be satisfied when $\lambda_{2D}N\geq 2\pi$. Close to this critical point, both widths become equal and tend to zero.

To obtain the value of the chemical potential we substitute the width obtained in Eq.~(\ref{2DSS}) into equation for $\dot{\phi}$ and get
\begin{equation}\label{beta2D}
\dot{\phi} \equiv - \mu_{2D} =
-\frac{\omega}{2\pi}\frac{2\pi^2-\lambda_{2D}^2N^2}{\sqrt{4\pi^2-\lambda_{2D}^2N^2}}=
-\frac{\omega}{2}\frac{1-2\eta^2}{\sqrt{1-\eta^2}},
\end{equation}
where $\eta=(\lambda_{2D}N)/(2\pi)$. In the analogy with 1D case we can write $\Psi (x,y,t)=e^{-i\mu_{2D} t}\Phi (x,y)$, Similar variational approach can be developed for the case of the quasi 1D square well potential. The only difference in the functional form of the effective Lagrangian, and therefore also in the equations of motion,  would be in potential term, which contains $\omega$,.
Finally we would like to point out that the analysis presented above can be used in the linear limit, when the waveguide mode structure can be predicted with satisfactory accuracy, see \cite{Li}.

\subsection{Collapse of the 2D wavefunction}\label{collaps}

We now consider the case when $\eta \geq 1$, i. e. in the regime where we expect the wavefunction to collapse. As we mentioned above, close to the collapse both widths of our solution, $W$ and $V$ become even. Hence, in the crude approximation, to describe the dynamics of the collapse we can assume axial symmetry. Upon neglecting harmonic potential contribution (which is negligible during the collapse) we obtain
\begin{equation}\label{colap}
\ddot{W}=\frac{1}{W^3}-\frac{\eta}{ W^3}=\frac{1-\eta}{W^3}.
\end{equation}
If we solve the Eq.~(\ref{colap}) with initial conditions $W(0)=W_{0}$, $\dot{W}(0)=0$, which correspond to the dynamics originated from some unstable state, 
we obtain
\begin{equation}
W(t)=W_0\sqrt{1-\frac{(\eta-1) t^2}{W_0^4}}
\end{equation}
The main conclusion from this simple calculation is that the collapse of the wavefunction occurs within the finite time (except when $\eta=1$, on the threshold for the collapse). The collapse time is equal to
\begin{equation}
t_{col}=\frac{W_0^2}{\sqrt{\eta-1}}.
\end{equation}
Collapse occurs on the timescale that is proportional to $W_0^2$, but what is more important it occurs the faster the the higher norm of the wavefunction is. This will be important in the next section when we discuss collapse during the soliton collision.

To relate our result to experiments, we calculate the critical atom number necessary to observe collapse in BEC. We consider a $^{85}$Rb condensate in a
highly anisotropic trap configuration with $\omega_z = 2\pi \times 350$ Hz, $\omega_y= 2\pi \times 55$ Hz, and a shallow confinement in the $x$ direction. This
configuration can be realized using an optical dipole trap \cite{trap}. For the scattering length value of $a_s= - 15 \,a_0$ \cite{collapse_exp2}
the critical number of atoms is $N_{\rm cr} \approx 10^3$, and the typical soliton dimensions close to the collapse threshold are of the order of several $\mu$m.

\section{Numerical results}

\subsection{Quasi-1D limit}

\begin{figure}[tbp]
\includegraphics[width=9cm]{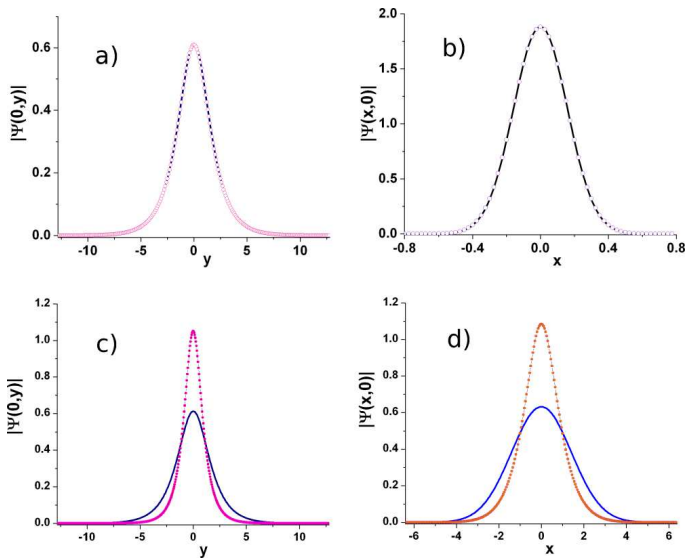}
\caption{Comparison of the 2D soliton cross sections (dots) (in $x$ and
$y$ planes) with quasi 1D approximation (continuous curves). a) and b)
corresponds to the frequency $\omega = 40$. We see a perfect agreement
between 1D soliton and cross section of 2D soliton along $y$ axis and
ground state of the harmonic potential and 2D soliton cross section along
$x$ axis. c) and d) were obtained for smaller frequency $\omega = 0.5$.
Some deviation from quasi 1D approximation can be observed.}\label{rys2}
\end{figure}

\begin{figure}[tbp]
\includegraphics[width=8cm]{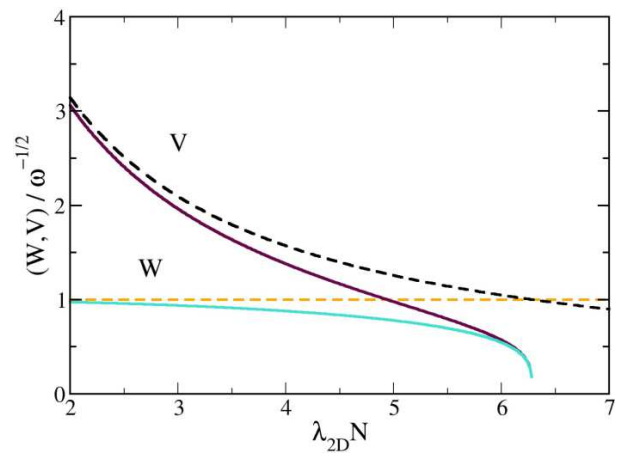}
\caption{Widths of the 2D soliton obtained from variational approximation, as a function of
parameter $\lambda_{2D}N$ (solid lines). Marked with dashed line are the corresponding values of
1D soliton width and the width of the ground state of the harmonic 1D potential (horizontal line).
}\label{rys3}
\end{figure}

\begin{figure}[tbp]
\includegraphics[width=8cm]{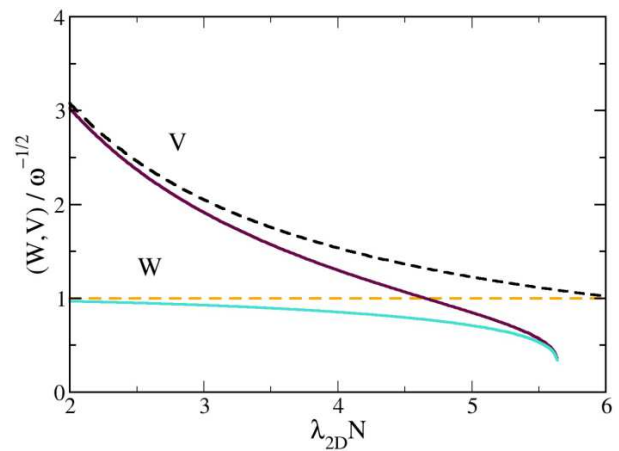}
\caption{Widths of the 2D soliton obtained from direct numerical simulations, as a function of
nonlinear parameter $\lambda_{2D}N$.}\label{rys4}
\end{figure}

\begin{figure}[tbp]
\includegraphics[width=8cm]{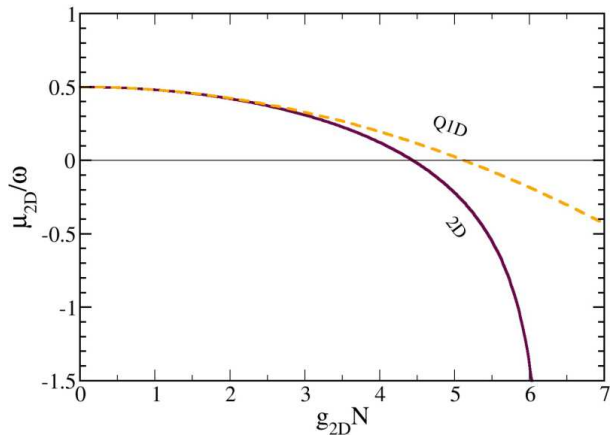}
\caption{Comparison of the chemical potential $\mu_{2D}$ obtained using quasi 1D approximation
(Eq.~(\ref{betaQ1D}) (dashed line) and full 2D variational approximation (Eq.~(\ref{beta2D})
(solid line) as a function of nonlinear parameter $\lambda_{2D}N$. Notice that the region between
the collapse and the point up to which quasi 1D approximation holds is very narrow.}\label{rys5}
\end{figure}

We now consider the case when value of $\lambda_{2D}$ (see Eq.~(\ref{NL2Dto1D})) tends to zero, but the value of $\lambda_{1D}$ and the norm remain constant. This corresponds to the limit $\omega \rightarrow \infty$. By substituting 2D nonlinear coefficient $\lambda_{2D}=\lambda_{1D}\sqrt{2\pi/\omega}$ into the expression for 2D eigenvalue $\mu_{2D}$ in Eq.~(\ref{beta2D}), we obtain
\begin{equation}\label{betaQ1D}
\mu_{2D}\simeq \frac{\omega}{2}-\frac{3\lambda_{1D}^2N^2}{8\pi} =
\frac{\omega}{2}+\mu_{1D}.
\end{equation}
This way we show that 2D eigenvalue, in the limit of high frequency, consists of 1D eigenvalue and the ground state energy of harmonic oscillator. It is a very intuitive result, since in this limit 2D soliton
becomes a product state of 1D soliton and the ground state of the harmonic potential in the transverse direction.  Now we examine the widths of 2D soliton presented in Eq.~(\ref{2DSS}). If we assume $\omega \rightarrow \infty $ one of the widths $W \simeq 1/\sqrt{\omega}$, and $V \rightarrow \sqrt{2\pi}/(\lambda_{1D}N)$. This is exactly the value we
obtained for the 1D soliton. In conclusion, in the limit considered here, both chemical potential and the width obtained from variational approximation take proper quasi 1D values. In Fig.~\ref{rys2} we present both cross-sections of 2D soliton in $x$ and $y$ planes. One of the cross sections is compared with 1D soliton and the other with the ground state of the harmonic potential. The comparison is made for two different values of frequency $\omega$. For larger value of $\omega$ we observe perfect matching and when $\omega$ becomes smaller, we can see slight deviations from quasi 1D approximation. In Fig.~\ref{rys3} we present 2D soliton width obtained from variational approximation, as a function of
parameter $\lambda_{2D}N$. For comparison we also included a dashed line representing a width of the 1D soliton, and straight dashes horizontal line corresponding to the width of the ground state of the trapping potential (Gaussian). In Fig.~\ref{rys4} the same comparison is made for the case of numerical solutions. In this case we defined the widths
numerically, as $(V,W)=\sqrt{\pi}(\langle|x|\rangle,\,\langle|y|\rangle)$, i.~e. the the mean value of the modulus of the coordinate in this direction. We clearly see that for small $\lambda_{2D}N$ 2D soliton width well approximates the equivalent 1D value, and when $\lambda_{2D}N$ approaches critical value, both widths become equal, while tending to zero. In this limit, the energy of the trapping potential is much smaller than kinetic and nonlinear energies, hence our solution becomes practically identical as that of Townes soliton \cite{Chiao}. Finally, in Fig.~\ref{rys5} we plot a chemical potential as a function of $\lambda_{2D}N$ obtained within quasi 1D approximation (Eq.~(\ref{betaQ1D})) and that obtained from full 2D variational analysis, (formula (\ref{beta2D})). Notice that when the $\mu_{2D}$ is of order of the excitation energy in the harmonic potential quasi 1D approximation breaks down. Also on this figure we can see a clear indication of the collapse region, when $\lambda_{2D}N$ approaches the value of $2\pi$, and the fact that region between the collapse and the point up to which quasi 1D approximation holds is very narrow.

\subsection{Soliton collisions}

Now we investigate collision of solitons that are moving along the quasi 1D harmonic potential presented in Fig.~\ref{potenfig}.
Soliton collisions in the one-dimensional NLSE were studied in many contexts, including nonintegrable dynamics of
vector solitons \cite{previous_collisions}.
In the case of BEC, the quasi-1D collisions were investigated in Ref. \cite{Salasnich} in the framework of the nonpolynomial Schr\"odinger equation (NPSE).
The results obtained here can be easily generalized for the quasi 1D square well potential.
The solitons that we used in the simulations were identical, and had equal and opposite velocities.

\begin{figure}[tbp]
\includegraphics[width=9cm]{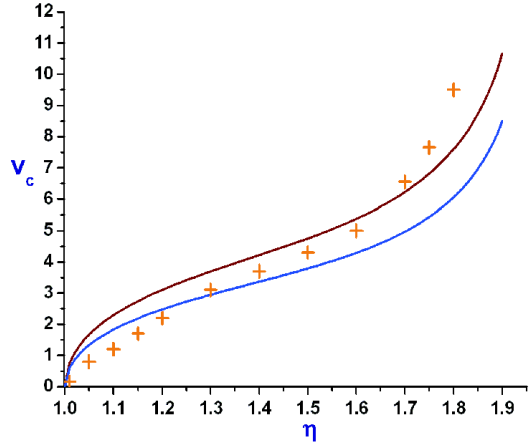}
\caption{Boundaries of the collisional collapse region for the trap with
the frequency $\omega=4$, obtained from variational approximation (with
Gaussian - upper continuous curve and hyperbolic secant trial functions -
lower continuous curve) and obtained from direct numerical simulations
(crosses). Solitons colliding with velocities below those marked with
crosses will experience collapse during their interaction.}
\label{figbond}
\end{figure}

We start with a pair of solitons, each of which is separately a solution of Eq.~(\ref{2DGP}). The initial width along free direction we denote by $V_{0}$ and the in the transverse direction by $W_{0}$. We assume that
the norm of each soliton as equal to $N/2$ and they travel with velocity $v$. The nonlinear interaction is described by the parameter $\lambda_{2D}$. We will introduce here a simple estimate of collapse during the collision based on variational approximation. It is well known that estimates based on variational analysis depends on the choice of variational basis, up to the multiplicative factor \cite{Anderson}. Hence, to improve the quality of our predictions we will consider the estimates obtained using Gaussian functions and hyperbolic secant functions. Lets first concentrate on Gaussians. The first parameter that describes collision is soliton interaction time. This is merely the time
when both wavefunctions have a significant overlap. If we assume that significant interaction appear as long as wavepackets are separated not more than FWHM, we obtain interaction time equal to
\begin{equation}
t_{int} \simeq \frac{2\gamma V_{0}}{v}=\frac{4\gamma W_0 }{\eta v},
\end{equation}
where $\gamma \simeq 1.178$. Notice that the width ratio differs from the one given in Eq.~(\ref{2DSS}), because soliton norm is now $N/2$ instead of $N$. Here we also assume that during this time interaction is almost
constant and is taken as that of full overlap (both function on top of each other). Now we turn to the condition for the collapse. As we see from Sec.~\ref{collaps} collapse is expected to occur for $\eta > 1$. This condition, in the case of collision, should refer to the situation when wavepackets fully overlap. On the other hand soliton with the norm
$N/2$ can only exist if $\eta < 2$. Taking both conditions into account we conclude that the region of interest is $1 < \eta < 2$. We estimate a time of collapse for solitons in the above interval to be
\begin{equation}
t_{col}=\frac{W_{0}^2}{\sqrt{\eta-1}}.
\end{equation}
Notice that here $W_0$ is the initial width of each of the colliding partners and given by
\begin{equation}
W_{0}=\sqrt[4]{\frac{4-\eta^2}{4\omega^2}}.
\end{equation}
From the above discussion it follows that there should be a critical velocity, above which during the collision the collapse will take place. This condition can be obtained by comparing time of collision and time of collapse
\begin{equation}
\frac{4\gamma W_{0}}{\eta v_{c}}\simeq \frac{W_{0}^2}{\sqrt{\eta - 1}},
\end{equation}
which after some algebra will give
\begin{equation}\label{vcgaus}
v_c(\eta) = \frac{4\gamma\sqrt{2\omega}}{\eta}\sqrt[4]{\frac{(\eta-1)^2}{4-\eta^2}}.
\end{equation}
If the velocity of solitons exceeds this critical value, there is not enough time during the collision to complete the collapse. If the velocity is lower than this value, during the overlap time collapse can fully develop. Hence the function $v_c(\eta)$ marks the boundary between regions of collapse and no collapse.

As we mentioned above, position of this line depends on the choice of the trial functions, which in the case just described were Gaussians. Had we chosen hyperbolic secant functions instead of Gaussians, we would have obtained slightly different result. First, we find that in this case the collapse for single soliton appear for the value of nonlinearity
($\lambda_{2D}N$) equal to 6 instead of $2\pi$. Hence, the parameter $\eta$ should be defined as $(\lambda_{2D}N)/6$, and after algebra, very similarly to what we presented for Gaussians, we obtain the condition for the critical velocity as
\begin{equation}\label{vcsec}
v_c(\eta) = \frac{8\gamma\sqrt{\omega}}{\sqrt{\pi}\eta}\sqrt[4]{\frac{(\eta-1)^2}{4-\eta^2}}.
\end{equation}
In Fig.~\ref{figbond} we show curves representing $v_c$ as a function of $\lambda_{2D}N$, obtained from Eqs.~(\ref{vcgaus}) and (\ref{vcsec}), indicated with solid lines, together with the results obtained from direct numerical simulations (crosses). The agreement between the variational predictions and direct numerical simulations is very good. It is somehow surprising. It was shown in Reference \cite{Garnier} that the variational analysis of the collapse dynamics does not work very well close to the threshold. Nevertheless it seems that for our crude estimate it is sufficient.
All the crosses are falling between curves obtained with Gaussian and hyperbolic secant trial functions. Solitons colliding with velocities below those marked with crosses will experience collapse during their interaction. In conclusion, we see that variational analysis predicts correctly and accurately the onset of catastrophic collapse during soliton collision.

\begin{figure}[tbp]
\includegraphics[width=7cm]{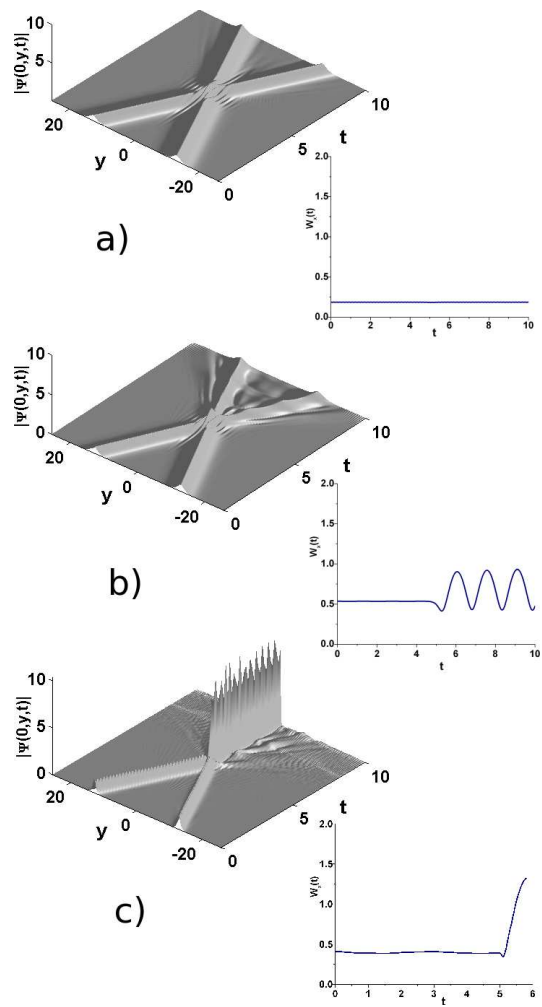}
\caption{Gallery of collisions. From quasi 1D to collapse. Soliton
velocity $v=2.5$. Panel a) high frequency ($\omega = 20$) and small
nonlinearity $\lambda_{2D}=2$ - solitons appear to be practically one
dimensional, and they asymptotically restore their original shape. Panel
b) small frequency $\omega = 2$ and nonlinearity the same as in previous
case - nonlinear interaction is comparable with excitation energy. We
observe oscillations with frequency $2\omega$ (inset on the right hand
side shows mean square radius in the transverse direction as a function
of time). Panel c) frequency the same as in case b), but higher
nonlinearity $\lambda_{2D}=3$ - we observe collapse during the
collision. } \label{figcol}
\end{figure}

The full picture of the soliton collision that emerges from our numerical studies is the following. Imagine we keep the norm of each soliton and their velocity constant. Then, depending on the strength of the confining potential ($\omega$ in the case of harmonic potential and the width in the case of quasi 1D square well potential) we can clearly identify three
regimes according to the relation between nonlinear interaction and the excitation energy. In the first regime, when the distance between ground state and first excited state of the confining potential is much bigger than the interaction energy, system is practically integrable, hence solitons pass though each other and restore their original shape after
the collision. When both energy scales become comparable we observe excitations in the transverse directions, corresponding to the transition between ground state and first coupled excited state. Finally, we observe catastrophic collapse during the collision. Behavior of solitons in these three regions is illustrated in Fig.~\ref{figcol}. Picture a) is
characteristic for the first region described above. It was obtained for high frequency $\omega = 20$ and small nonlinearity $\lambda_{2D}=2$. Solitons appear to be practically one dimensional, and they asymptotically (after the collision) restore their original shape. In the case b) $\omega = 2$ and nonlinearity is the same as in previous case and
nonlinear interaction becomes comparable with excitation energy. Some of the population will be transferred to the excited state and we observe the beats of the frequency $2\omega$ in the transverse direction. It can be detected for example by calculating mean square radius in the transverse direction, shown in the inset on the right hand side. Finally, in picture c) where $\lambda_{2D}=3$ and $\omega = 2$, total population is above the critical value and collision time is sufficiently long - we observe collapse during the collision. Analogous effect was previously observed in optics, when two spatial solitons collided in planar waveguides, see for instance \cite{Buttner}.

%

\section{Conclusions}

In conclusion, we analyzed the stability and collisions of quasi 1D solitons in the confining potential, both numerically and within variational approximation. We showed that variational approximation constitutes an excellent basis for estimating physical parameters of quasi 1D solitons. The advantage of using this approximation is that one obtains analytical formulas, and can predict soliton behavior in different limits and under various circumstances. One of the examples is
the threshold of the collapse and its dynamics. Comparing time of collapse and collision time we could predict the result of the soliton collision and identify three different regimes according to the relation between nonlinear interaction and the excitation energy. In the first regime, when the distance between ground state and first excited state of the confining potential is much bigger than the nonlinear interaction, system is practically integrable, hence solitons pass through each other and restore their original shape after the collision. When both energy scales become comparable we observe excitations in the transverse directions, corresponding to the transition between ground state and first coupled excited state. Finally we observe catastrophic collapse during the collision. In the following paper we will present similar considerations for the case of Gap solitons.

\section{Acknowledgements}

M.T. would like to thank Prof. Yuri Kivshar for the valuable discussions and hospitality during his stay at ANU and acknowledges the support of the Polish Government Research Grant for 2006-2009.  N. V. H. was supported by Polish Ministry of Science and Education under grant N202 014 31. M.M. acknowledges support from the Foundation for Polish Science and ARC Center of Excellence for Quantum Atom Optics.


\begin{thebibliography}{99}

%
%
%
%
%
%
%
%
%
%
%
%
%
%
%
%
%
%
%
%
%
%
%
%
%
%
%

\bibitem{Zahar} V. E. Zakharov and A. B. Shabat,
Sov. Phys. JETP {\bf 34}, 62
(1972).

\bibitem{Dimitriev} S. V. Dmitriev, D. A. Semagin, A. A.
Sukhorukov, and T. Shigenari1,
Phys. Rev. E {\bf 66}, 046609 (2002).

\bibitem{Gardiner} A. D. Martin, C. S. Adams, and S. A. Gardiner,
Phys. Rev. Lett. {\bf 98},
020402 (2007).

\bibitem{SSSience} G. I. Stegeman and M. Segev, Science {\bf 286},
1518 (1999).

\bibitem{ALPapier} D. Andersen and M. Lisak, Phys. Rev. A {\bf 32}, 2270
(1995).

\bibitem{Ciamaro} P. Chamorro-Posada and G. S. McDonald, Phys Rev.
E {\bf 74}, 036609 (2006).

\bibitem{Hector} H. E. Nistazakis, D. J. Frantzeskakis and B. A. Malomed,
Phys. Rev. E, {\bf 64}, 026604 (2001).

\bibitem{BECexp} L. Khaykovich, F. Schreck, G. Ferrari, T. Bourdel, J. Cubizolles, L. D. Carr, Y. Castin, and C. Salomon, Science {\bf 296}, 1290 (2002); K. E. Strecker, G. B. Partridge, A. G. Truscott and R. G. Hulet, Nature {\bf 417}, 150 (2002); V. M. P\'erez-Garc\'ia, H. Michinel, and H. Herrero, Phys. Rev. A {\bf 57}, 3837 (1998).

\bibitem{Solzat} L. D. Carr, J. Brand, Phys. Rev. A {\bf 70}, 033607 (2004); M. I. Rodas-Verde, H. Michinel, V. M. Perez-Garcia, Phys. Rev. Lett. {\bf 95}, 153903 (2005); A. V. Carpentier, H. Michinel, Europhys. Lett {\bf 78}, 10002 (2007).

\bibitem{Salasnich} L. Salasnich, A. Parola, and L. Reatto, Phys.
Rev. A {\bf 65}, 043614 (2001); ibid Phys. Rev. A  {\bf 66}, 043603 (2002).

\bibitem{collapse_exp} J. M. Gerton, D. Strekalov, I. Prodan, and R. G. Hulet, Nature {\bf 408}, 692 (2000).

\bibitem{collapse_exp2} E. A. Donley, N. R. Claussen, S. L. Cornish, J. L. Roberts, E. A. Cornell, and C. E. Wieman, Nature {\bf 412}, 295 (2001).

\bibitem{Li} Q. Y. Li, C. Pask, R. A. Sammut, Opt. Lett. {\bf 16},
1083 (1991); R. A. Sammut and C. Pask,  J. Opt. Soc. Am. B {\bf 8}, 395
(1991).

\bibitem{Obert} K.M. Hilligs{\o}e, M.K. Oberthaler, and K.-P.
Marzlin, Phys. Rev. A {\bf 66}, 063605 (2002).

\bibitem{Kajak} L. Khaykovich and B. A. Malomed, Phys. Rev. A
{\bf 74}, 023607 (2006).

\bibitem{Buttner} O. Büttner, M. Bauer, S. O. Demokritov,
B. Hillebrands, M. P. Kostylev, B. A. Kalinikos A. N. Slavin, Phys. Rev. Lett.
{\bf 82}, 4320 (1999).

\bibitem{Raghavan} S. Raghavan, G. P. Agrawal, Opt. Commun.
{\bf 180}, 377 (2000).

\bibitem{Chiao} R. Y. Chiao, E. Germire and C. H. Townes, Phys. Rev. Lett {\bf 13}, 479
(1964).

\bibitem{Progres} B.~A.~Malomed, in: Progress in Optics, vol. {\bf 43}, p. 71
(ed. by E. Wolf:\ North Holland, Amsterdam, 2002).

\bibitem{VABEC} V. M. P\'erez-Garc\'ia, H. Michinel, J. I. Cirac, M. Lewenstein, and P. Zoller, Phys. Rev. Lett. {\bf 77}, 5320 (1996).

\bibitem{trap} L. E. Sadler, J. M. Higbie, S. R. Leslie, M. Vengalattore, and D. M. Stamper-Kurn, Nature {\bf 443}, 312 (2006).

\bibitem{previous_collisions} B. A. Malomed and S. Wabnitz, Opt. Lett. {\bf 16}, 1388 (1991);
G. Huang, M. G. Velarde, and V. A. Makarov, Phys. Rev. A {\bf 64}, 013617 (2001);
J. Babarro, M. J. Paz-Alonso, H. Michinel, J. R. Salgueiro, and D. N. Olivieri, Phys. Rev. A {\bf 71}, 043608 (2005).

\bibitem{Anderson} M. Desaix, D. Anderson, M. Lisak, J. Opt. Soc. Am. B {\bf 8}, 2082 (1991).

\bibitem{Garnier} J. Garnier, F. Kh. Abdullaev, B. B. Baizakov, Phys. Rev. A {\bf 69},053607 (2004).



\end{thebibliography}
\end{document}